\documentclass[
 reprint,
superscriptaddress,
 amsmath,amssymb,
 aps,
pra,
nofootinbib,longbibliography]{revtex4-1}
\usepackage{hyperref}
\usepackage{graphicx}
\usepackage{dcolumn}
\usepackage{subcaption}
\usepackage{bm}
\usepackage{verbatim}
\usepackage{float}
\usepackage{balance}
\usepackage{natbib}
\usepackage{color}
\usepackage{hyperref}
\usepackage{graphicx}
\usepackage{dcolumn}
\usepackage{bm}
\usepackage{soul}

\newcommand{\bi}{\mathbf}

\begin{document}
\title{Robustness of quantized Hall resistivity under cavity coupling at zero temperature}

\author{Juan~Román-Roche}
\affiliation {Instituto de Nanociencia y Materiales de Aragón (INMA), CSIC-Universidad de Zaragoza, Zaragoza 50009, Spain}
\affiliation{Departamento de Física de la Materia Condensada, Universidad de Zaragoza, Zaragoza 50009, Spain}

\author{Jie~Wang}
 \affiliation{International Center for Quantum Materials, Peking University, Beijing 100871, China}
\affiliation{Beijing Key Laboratory of Quantum Devices, Peking University, Beijing 100871, China}

\author{Michael~Ruggenthaler}
\affiliation{Max Planck Institute for the Structure and Dynamics of Matter, Hamburg, Germany}

\author{Angel~Rubio}
\affiliation{Max Planck Institute for the Structure and Dynamics of Matter, Hamburg, Germany}
\affiliation{Initiative for Computational Catalysis (IIC) and Center for Computational Quantum
Physics (CCQ), The Flatiron Institute, 162 Fifth Avenue, New York, NY 10010,
USA}

\author{Vasil~Rokaj}
 \email{vasil.rokaj@villanova.edu}
\affiliation{Department of Physics, Villanova University, 
             Villanova, Pennsylvania 19085, USA}

\begin{abstract}
Recent experiments have shown that strong light–matter coupling in electromagnetic cavities can modify transport properties of quantum Hall systems through the formation of Landau polaritons, prompting questions about the robustness of topological protection. While earlier theory demonstrated that the Hall conductivity can be modified at finite temperature and finite polariton lifetime (or finite broadening), experiments primarily probe the resistivity tensor. Our phenomenological model reveals an asymmetry between conductivity and resistivity in quantum Hall systems under strong light–matter interaction, showing that at zero temperature the Hall resistivity remains completely immune to cavity-induced modifications arising from polariton broadening, independent of the light–matter coupling strength. These results provide a deeper explanation for the absence of renormalization in the von Klitzing constant in experiments probing the even QH plateaus through the Hall resistivity at low temperature, and clarify the distinct roles of dissipation and strong light–matter coupling in hybrid light–matter systems.
\end{abstract}

\maketitle

\section{Introduction}

In recent years, a convergence between condensed matter physics and quantum optics, driven by advances that enable quantum materials to be manipulated with cavity fields, has given rise to cavity quantum materials, ~\cite{SchlawinSentefReview, BaydinCavityReview, VidalEbbesenReview, RuggiReview2023, ITeReview}. Confining electromagnetic fields in an optical cavity substantially enhances the light–matter coupling between electrons in the material and cavity photons, and experiments suggest that the equilibrium properties of quantum materials can be modified by the enhanced vacuum fluctuations of the electromagnetic field inside an optical cavity~\cite{SchlawinSentefReview, BaydinCavityReview, VidalEbbesenReview,  ITeReview}. These striking experimental demonstrations include cavity-modified transport in semiconductors due to exciton-polariton formation~\cite{Delor2023}, cavity control of a metal-to-insulator phase transition~\cite{FaustiThermal2023}, and remarkably cavity-altered superconductivity was reported recently due vibrational coupling to infrared hyperbolic modes~\cite{CavityAlteredSuperconductivity}, and via vibrational strong coupling to surface plasmon polaritons~\cite{EbbesenSuperconductivity}. In addition, significant experimental effort has been put forward to design novel chiral cavities which break time-reversal symmetry~\cite{suarez2024chiral,PhysRevB.109.L161302,Tay2024,kulkarni2025realizationchiralphotoniccrystalcavity}.

Moreover, experiments on quantum Hall (QH) systems embedded in terahertz cavities have demonstrated that cavity-mediated interactions can modify quantum transport, optical response, and electronic correlations in both the integer and fractional QH regimes~\cite{ScalariScience, li2018, paravacini2019, FaistCavityHall, enkner2024enhancedfractional}. These striking phenomena are typically associated with the observation of ultrastrong light-matter interaction and the emergence of hybrid quasiparticles known as Landau polaritons, formed from Landau levels mixed with cavity photons. Landau polaritons and the interactions they mediate have been linked to measurable changes in magnetotransport, including modifications of Shubnikov–de Haas oscillations~\cite{paravacini2019}, enhancement of fractional quantum Hall gaps~\cite{enkner2024enhancedfractional}, and a breakdown of the topological protection in integer QH states~\cite{FaistCavityHall}. This unexpected result that has sparked significant theoretical interest, and remains not fully understood. The experimental breakdown was firstly argued to originate from a disorder-assisted cavity-mediated long-range hopping~\cite{CiutiHopping}, and subsequently an alternative analytical Landau polariton model~\cite{rokaj2023topological} suggested the weakening of the topological protection in the IQH \cite{FaistCavityHall} due to a lowering of the thermal activation gap, caused by a lower polariton mode that is softer than the bare Kohn mode~\cite{CyclotronKohnTheorem}. 

The weakened topological protection in Ref.~\cite{rokaj2023topological} appeared in the components of the Hall conductivity tensor and was the result of three factors, (i) the ultrastrong light-matter interaction, (ii) finite-temperature, and (iii) a finite polariton lifetime modeled by a broadening parameter in the polariton linewidths. The cavity modification of the Hall conductance persists even at zero temperature as long as the effective polariton broadening is not zero. At zero broadening and zero temperature, the Hall conductance recovered its expected quantized value for any value of the light-matter coupling strength, in agreement with the theorems about the topological protection of QH states~\cite{ThoulessNiuQuantization, LaughlinPRB, Thouless, QHGirvinLectures}. In most of cavity QH experiments however, the resistivity tensor is the one that is probed~\cite{FaistCavityHall, paravacini2019, enkner2024enhancedfractional}. Importantly, in Hall systems conductivity and resistivity encode distinct physical information, as they are related by tensor inversion rather than simple scalar reciprocity. 

In this paper, we revisit the Landau polariton model in Ref.~\cite{rokaj2023topological} aiming to examine the behavior of the resistivity tensor at zero temperature, where the theorems about the topological protection of the QH systems are formulated and strictly apply~\cite{LaughlinPRB, Thouless, QHGirvinLectures}. We find that the Hall resistivity is totally immune to any modification coming from the finite polariton lifetime independently of the the light-matter interaction strength. This is in stark contrast to the Hall conductivity which bears an effect from the combination of finite polariton broadening and light-matter interaction. This is a remarkable finding as it suggests that there is a fundamental asymmetry on the response of Hall conductivity and resistivity under strong light-matter interaction. This finding provides a deeper explanation for the absence of any modification or renormalization in the von Klitzing constant in recent experiments probing the even integer QH plateaus via measurements of the Hall resistivity at low temperature~\cite{RokajPRX}. However, it is important to note that cavity-induced temperature activation effects have been observed~\cite{FaistCavityHall, RokajPRX, enkner2024enhancedfractional} and thus the combined effects of temperature, finite polariton lifetime and ultrastrong light-matter interaction are still relevant~\cite{rokaj2023topological}.

In addition, we identify that there is an anisotropic behavior in the resistivity tensor between the longitudinal and the Hall resistivity, which originates from the preferred polarization direction of the cavity photon field, as we find that the longitudinal resistivity is affected by the cavity. Finally, we discuss and compare the behavior of both conductivity and resistivity in the zero frequency limit for the cavity photons where the lower polarion gap closes. As found in previous works~\cite{rokaj2023topological, RokajButterfly2022} the Hall conductivity is modified when the lower polariton gap closes, even for a zero polariton broadening, while the Hall resistivity still remains unaffected.

\section{Model Hamiltonian for two-dimensional electron gas}
\label{sec:model}

Our model considers a two-dimensional electron gas coupled to a strong magnetic field and a single-mode homogeneous cavity field. The system is described by the Pauli-Fierz Hamiltonian~\cite{spohn2004, cohen1997photons, rokaj2017}
\begin{equation}
H = \sum^{N}_{i=1}\frac{\big(\bm{\pi}_i+e\bi{A}\big)^2}{2m} + \hbar\omega\left(a^{\dagger}a + \frac{1}{2}\right) + \sum_{i<j}W(\bi{r}_i-\bi{r}_j),\label{Pauli-Fierz}
\end{equation}
where $\bm{\pi}_i=\textrm{i}\hbar\nabla_i+e\bi{A}_{\textrm{ext}}(\bi{r}_i)$ are the dynamical momenta of the electrons, and $\mathbf{A}_{\textrm{ext}}(\mathbf{r})=-\bi{e}_xBy$ describes the applied magnetic field $\bi{B}=\nabla \times \bi{A}_{\textrm{ext}}(\bi{r})=B\bi{e}_z$. The cavity field $\bi{A} = \sqrt{\frac{\hbar}{2\epsilon_0\mathcal{V}\omega}}\bi{e}_x\left(a + a^{\dagger}\right)$ is characterized by the in-plane polarization vector $\bi{e}_x$ and the photon's bare frequency $\omega$. The $\mathcal{V}$ and $\epsilon_0$ are the effective mode volume and the dielectric constant, respectively. The operators $a$ and $a^{\dagger}$ represent photonic annihilation and creation operators which satisfy bosonic commutation relations $[a,a^{\dagger}]=1$. Further, $W(\bi{r}_i-\bi{r}_j)=1/4\pi \epsilon_0 |\bi{r}_i-\bi{r}_j|$ is the Coulomb interaction between the electrons. We have parameterized the bare electron dispersion by an effective mass $m$ and assumed Galilean invariance. With Galilean invariance in a homogeneous system, the center-of-mass (CM) is decoupled from the relative motion of the electrons, regardless of the interaction strength~\cite{CyclotronKohnTheorem}. The kinetics of the CM and its coupling to light is best described in terms of the scaled CM coordinate $\mathbf{R}=(X,Y)=\sum^N_{i=1}\mathbf{r}_i/\sqrt{N}$ where $N$ is the total particle number. Following the derivation presented in Appendix~\ref{Polariton Solution} we obtain the Hamiltonian describing the coupling of the CM to light 
\begin{equation}
H_{\textrm{cm}}=\frac{1}{2m}\left(\bm{\Pi}+e\sqrt{N}\mathbf{A}\right)^2+\hbar\omega\left(a^{\dagger}a+\frac{1}{2}\right)    
\end{equation}
where $\bm{\Pi}=\textrm{i}\hbar \nabla_{\mathbf{R}}+e\mathbf{A}_{\textrm{ext}}(\mathbf{R})$ is the dynamical momentum of the CM. It is important to mention that if we break Galilean invariance or consider a spatially inhomogeneous cavity field, the relative degrees of freedom will couple to quantum light. The CM Hamiltonian has the form of two coupled harmonic oscillators, one for the Landau level transition and one for the photons. Such a Hamiltonian is known as the Hopfield Hamiltonian which can be solved by the Hopfield transformation~\cite{Hopfieldmodel}. The Hopfield model has been employed in previous works for the description of single-particle Landau level transitions coupled to cavity photons~\cite{Hagenmuller2010cyclotron, BartoloCiuti}. Here, it shows up for the collective coupling of the electrons which emerges naturally through the CM. After the Hopfield transformation we find
\begin{equation}
   H_{\rm cm} = \hbar\Omega_+\left(b^\dag_+b_++\frac12\right) + \hbar \Omega_-\left(b^\dag_-b_-+\frac12\right)
\end{equation}
where $\{b^{\dagger}_{\pm},b_{\pm}\}$ are the creation and annihilation operators of the Landau polariton quasiparticles satisfying bosonic commutation relations $[b_l,b^{\dagger}_{l^{\prime}}]=\delta_{ll^{\prime}}$ with $l,l^{\prime}=\pm$. The details about the diagonalization of $H_{\rm{cm}}$ are given in the SM. The $\Omega_{\pm}$ are the upper and lower Landau polariton modes respectively,
\begin{eqnarray}\label{Polariton modes}
 \Omega^2_{\pm}=\frac{\omega^2+\omega^2_d+\omega^2_c}{2}\pm\sqrt{\omega^2_d\omega^2_c+\left(\frac{\omega^2+\omega^2_d-\omega^2_c}{2}\right)^2}
\end{eqnarray} 
where $\omega_d=\sqrt{e^2N/m\epsilon_0 \mathcal{V}}$ is the diamagnetic frequency originating from the $\mathbf{A}^2$ term which depends on the number of electrons $N$ and the effective mode volume $\mathcal{V}$, and $\omega_c=eB/m$ is the cyclotron frequency~\cite{Landau}. To define the polariton operators we represent $\{a,a^{\dagger}\}$ in terms of a displacement coordinate $q$ and its conjugate momentum $\partial_q$ as $a = (q+\partial_q)/\sqrt2$ with $a^{\dagger}$ obtained via conjugation~\cite{cohen1997photons, spohn2004}. The polariton operators then are written in terms of mixed coordinates as $S_{\pm}=\sqrt{\hbar/2\Omega_{\pm}}\left(b_{\pm}+b^{\dagger}_{\pm}\right)$ with
\begin{eqnarray}
    S_+ = \frac{\sqrt{m} \bar Y+q\Lambda\sqrt{\hbar/\omega}}{\sqrt{1+\Lambda^2}}\; \textrm{and}\; S_- = \frac{-q\sqrt{\hbar/\omega} +\sqrt{m} \Lambda \bar Y}{\sqrt{1+\Lambda^2}}\nonumber
\end{eqnarray}
where $\bar Y=Y+\frac{\hbar K_x}{eB}$ is the guiding center and $K_x$ is the electronic wave number in the $x$-direction. Also we introduced the parameter $\Lambda=\alpha-\sqrt{1+\alpha^2}$ with $\alpha=\left(\omega^2_c-\omega^2-\omega^2_d\right)/2\omega_d\omega_c$ which quantifies the mixing between electrons and photons.

\section{Quantum Hall Resistivity in a Cavity}\label{sec:transport}

\subsection{General case}

The gauge-invariant current operator for homogeneous fields solely depends on the CM dynamical momentum and the cavity field~\cite{Landau, Rokaj2022} $\mathbf{J}=-\frac{e\sqrt{N}}{m}\left(\mathbf{\Pi}+e\sqrt{N}\mathbf{A}\right)$ (see also Appendix~\ref{Transport}). Due to this property and the separability of $H_{\textrm{cm}}$ from the electronic correlations we can compute the transport of the system by focusing only on the states of $H_{\rm cm}$. At $T=0$ the system is in the polariton vacuum $|\Psi_{\textrm{gs}}\rangle=|0_+\rangle|0_-\rangle$ which is annihilated by both polariton operators $b_{\pm}$. Given this state, we employ the standard Kubo formalism~\cite{kubo} for the computation of the current correlators $\chi_{ab}(t)=-\textrm{i}\Theta(t) \langle \Psi_{\rm gs}|[J_{a}(t),J_{b}]|\Psi_{\rm gs}\rangle/\hbar$ in the time domain which we transform to the frequency domain in order to obtain the optical conductivities~\cite{kubo} $\sigma_{ab}(w)=\frac{\textrm{i}}{w+\textrm{i}\delta}\left(\frac{e^2n_{2d}} {m}\delta_{ab}+\frac{\chi_{ab}(w)}{A}\right)$ where $A$ and $n_{2d}=N/A$ are the area and the electron density of the 2d material respectively, $\delta$ is the broadening parameter, and $\delta_{ab}$ the Kronecker delta with $a,b \in \{x,y\}$. The optical conductivities $\sigma_{ab}(w)$ are given in the frequency domain in terms of the frequency $w$. The full details for the transport computations are provided in Appendix~\ref{Transport}. The poles of the response functions $\chi_{ab}(w)$ identify the optical responses of the system and its excitations. In addition, using the Kubo formula we find the Hall and longitudinal DC ($w=0$) conductivities \cite{rokaj2023topological}, which can be written as
\begin{eqnarray}
&&\sigma_{xy}= - \sigma_{yx} = \frac{e^2\nu}{h(1+\Lambda^2)}\left[\Delta_+ +\Lambda^2\Delta_-\right]\\
    &&\sigma_{xx}= \sigma_D\left[1-\frac{1}{1+\Lambda^2}\left(\lambda_+ \Delta_+ +\Lambda^2 \lambda_- \Delta_-\right)\right]\\
    &&\sigma_{yy}=\sigma_D\left[1-\frac{1}{1+\Lambda^2} \left(\Delta_+ +\Lambda^2\Delta_-\right)\right]
\end{eqnarray}
with
\begin{eqnarray}
\lambda_{\pm}=\frac{\Omega^2_{\pm}}{\omega^2_c} \;\; \textrm{and}\;\; \Delta_{\pm}=\frac{\Omega^2_{\pm}}{\Omega^2_{\pm}+\delta^2} .
\end{eqnarray}
Note that $\sigma_{D}=e^2n_{2d}/m\delta$ is the Drude DC conductivity, and that in $\sigma_{xy}$ we introduced the Landau level filling factor $\nu=n_{2d}h/eB$~\cite{Peierls, Tong}.
As established in Ref. \cite{rokaj2023topological}, all components of the conductivity tensor, and in particular the transverse conductivity, exhibit deviations with respect to their quantized value in the absence of a cavity, as long as a non-zero polariton broadening $\delta \neq 0$ is kept. Notably, the effect for the transverse component vanishes in the zero dissipation limit: $\lim_{\delta \to 0} \sigma_{xy} = e^2 \nu / h$, which is in agreement with the theorems about the topological protection of QH states~\cite{ThoulessNiuQuantization, LaughlinPRB, Thouless, QHGirvinLectures}.

In this paper, we clarify the combined effect of the cavity coupling and the finite polariton lifetime, on the transport properties of the electron gas, by additionally analyzing the resistivity tensor. To obtain the resistivity tensor
\begin{equation}
\rho=\sigma^{-1}=\frac{1}{\sigma_{xx}\sigma_{yy}+\sigma_{xy}^2}\left(\begin{matrix}
\sigma_{yy} & -\sigma_{xy} \\
\sigma_{xy} & \sigma_{xx} 
\end{matrix} \right)=\left(\begin{matrix}
\rho_{xx} & \rho_{xy} \\
\rho_{yx} & \rho_{yy} 
\end{matrix} \right),
\end{equation}
it is convenient to establish the following identities
\begin{eqnarray}\label{lambda delta1}
1+\Lambda^2=\lambda_++\Lambda^2\lambda_-\;\; \textrm{and}\;\; \lambda_{\pm}(1-\Delta_{\pm})=\frac{\delta^2}{\omega^2_c
}\Delta_{\pm}
\end{eqnarray}
which can be deduced directly from the definition of the mixing parameter $\Lambda$. From Eq.(\ref{lambda delta1}) we can derive another useful relation,
\begin{eqnarray}\label{the crucial one}
    \lambda_+ \Delta_++\Lambda^2\lambda_-\Delta_-=1+\Lambda^2-\frac{\delta^2}{\omega^2_c}\left(\Delta_++\Lambda^2\Delta_-\right).
\end{eqnarray}
Using the above relation $\sigma_{xx}$ can be written as follows,
\begin{eqnarray}
    \sigma_{xx}=\frac{\sigma_D}{1+\Lambda^2}\frac{\delta^2}{\omega^2_c}\left(\Delta_++\Lambda^2\Delta_-\right)
\end{eqnarray}
We can now compute the determinant of the conductivity tensor. First, we have
\begin{eqnarray}
\sigma_{xx}\sigma_{yy}=\frac{(e^2 \nu/h)^2}{(1+\Lambda^2)^2}\left[\Delta_++\Lambda^2\Delta_-\right]\left[1+\Lambda^2(1-\Delta_-)- \Delta_+\right]\nonumber
\end{eqnarray}
where we have used that the Drude weight can be written in terms of the filling factor as follows $\sigma_D=(\nu e^2/h)(\omega_c/\delta)$. 
Adding to the above the quadrature of the Hall conductivity $\sigma^2_{xy}$,
\begin{eqnarray}
    \sigma^2_{xy}=\frac{(e^2\nu/h)^2}{(1+\Lambda^2)^2}\left(\Delta_++\Lambda^2\Delta_-\right)^2,
\end{eqnarray}
we complete the determinant of $\sigma$,
\begin{eqnarray}
    \sigma_{xx}\sigma_{yy}+ \sigma^2_{xy}=\frac{(e^2\nu/h)^2}{(1+\Lambda^2)} \left(\Delta_++\Lambda^2\Delta_-\right).
\end{eqnarray}
Using the above result we obtain the final expressions for the components of the resistivity tensor.
\begin{eqnarray}\label{quantized}
    &&\rho_{yx}=\frac{h}{e^2\nu} \label{eq:hallresistivity}\\
    &&\rho_{xx}=\sigma_{\rm D}^{-1} \frac{\omega^2 + \omega_{\rm d}^2 + \delta^2}{\omega^2 + \delta^2} \\
    &&\rho_{yy}=\sigma_{\rm D}^{-1}
\end{eqnarray}
For $\rho_{xy}$ we recover the quantized value for the Hall resistivity without any effect from the light-matter interaction or the broadening parameter.
Thus, in contrast to the Hall conductivity $\sigma_{xy}$ which for finite broadening $\delta$ is affected by the light-matter coupling, the Hall resistivity $\rho_{yx}$ remains unaffected due its dissipationless nature.
Likewise, for $\rho_{yy}$ we recover the Drude conductivity.
We see now that the modifications of all the conductivity components arise solely from a modification to the longitudinal resistivity in the direction of the cavity field $\rho_{xx}$. It is also interesting to note that $\rho_{xx}$ is independent of the cyclotron frequency. 
The modification that the cavity induces in this longitudinal resistivity component is not affected by the presence of the classical transverse magnetic field. 
In fact, this is the same modification that was obtained in the 2d electron gas in a cavity without classical magnetic field in Ref. \cite{Rokaj2022}.
The only effect of the classical field is to induce
the typical Hall physics that give rise to non-zero transverse resistivities and
conductivities. 
There are no additional effects from the simultaneous presence
of the classical magnetic field and the cavity field.

\subsection{The zero cavity frequency limit}

A particularly interesting point of this model is that the zero dissipation $\delta \to 0$ and zero cavity frequency $\omega \to 0$ limits are not commutative. 
This is due to the fact that the model exhibits singular behavior in the zero cavity frequency limit, since the lower polariton $\Omega_-$ goes to zero and there is a gap closing. Topological protection is expected to be broken when there is a gap closing. In our phenomenological model this indeed occurs if the cavity frequency is taken to zero, and this cavity induced phenomenon survives even after removing any dissipation effects by taking the broadening to zero,
\begin{equation}
    \lim_{\delta \to 0} \left( \lim_{\omega \to 0} \sigma_{xy}\right)= \frac{e^2\nu}{h} \frac{1}{1 + \Lambda^2} ,
\label{eq:Hallconductivitynodissip}
\end{equation}
with $\Lambda = \omega_{\rm d} / \omega_{\rm c}$ in this limit. This is in agreement with the result of Ref. \cite{RokajButterfly2022} which studied transport in the absence of dissipation (no broadening) and in the $\omega \to 0$ limit. On the other hand, the opposite order of limits results in no modification of the Hall conductivity
\begin{equation}
    \lim_{\omega \to 0} \left( \lim_{\delta \to 0} \sigma_{xy}\right)= \frac{e^2\nu}{h}.
\end{equation}

Nevertheless, we show here that the non-dissipative modification of the Hall conductivity of Eq. \eqref{eq:Hallconductivitynodissip}, due to the lower polariton gap closing $\Omega_- \to 0$, still does not result into a modification of the Hall resistivity. As shown in Eq. \eqref{eq:hallresistivity}, the quantized value of the Hall resistivity is independent of $\omega$ and $\delta$. As such, it remains unmodified in the zero cavity frequency limit $\omega \to 0$, and its value is independent on the ordering of the $\delta \to 0$ and $\omega \to 0$ limits. In the resistivity tensor, all the dependence on $\omega$ and $\delta$ appears in the longitudinal component $\rho_{xx}$, which also exhibits the non-commutativity of limits that is observed in the conductivities.

\section{Discussion and Outlook}\label{sec:discussion}

Our phenomenological model establishes a clear and physically transparent distinction between the cavity response of the Hall conductivity and the Hall resistivity in quantum Hall systems coupled to cavity photons. While previous work~\cite{rokaj2023topological} demonstrated that the Hall conductivity can deviate from its quantized value due to the combined effects of ultrastrong light–matter interaction, finite polariton lifetime, and finite temperature, we show that the Hall resistivity remains strictly immune to polariton broadening at zero temperature. This finding clarifies that topological robustness in cavity QH systems manifests itself, independently of light-matter interaction strength and polariton broadening, at the level of the resistivity tensor . This asymmetry is particularly relevant in light of recent experiments, where transport measurements are naturally performed in terms of resistivity~\cite{FaistCavityHall, RokajPRX}. Our results provide a deeper explanation for why no renormalization of the von Klitzing constant in the even integer QH plateaus has been observed in high-precision measurements at low temperature under ultrastrong coupling~\cite{RokajPRX}. In this sense, the Hall resistivity emerges as the correct observable for assessing topological protection in driven or hybrid QH platforms. We would like to mention that recently a cavity QH hydrodynamical approach~\cite{cardoso2026CQH} was developed which reaches similar conclusions with this work.

Beyond topological robustness, we identify a qualitative anisotropy in the resistivity tensor: while the Hall resistivity remains unaffected, the longitudinal resistivity acquires cavity-induced corrections that explicitly depend on the polarization direction of the cavity field. This highlights how cavity photons can selectively break rotational symmetry at the level of electronic transport, even when topological invariants remain intact. Such anisotropies may provide experimentally accessible signatures of cavity-mediated interactions beyond Hall plateaus.
Our analysis also sheds light on the singular zero-frequency limit associated with the closing of the lower polariton gap. In agreement with earlier studies~\cite{RokajButterfly2022}, we find that the Hall conductivity is sensitive to this limit even in the absence of polariton broadening, whereas the Hall resistivity retains its quantized value. This further reinforces the notion that conductivity and resistivity encode fundamentally different physical information in hybrid light–matter systems.
Looking ahead, several open directions remain. Incorporating finite temperature effects in a fully self-consistent manner, including realistic disorder and dissipation mechanisms, will be essential to bridge zero-temperature topological arguments with experimentally observed activated transport. Extending our framework to fractional quantum Hall states and to chiral cavities that explicitly break time-reversal symmetry~\cite{suarez2024chiral,PhysRevB.109.L161302,Tay2024,kulkarni2025realizationchiralphotoniccrystalcavity} may also reveal new regimes where cavity photons reshape correlations without compromising topological protection. More broadly, our work underscores that cavity quantum materials demand a careful reassessment of which observables faithfully reflect topological robustness and how light–matter hybridization reshapes fundamental transport properties.

\begin{acknowledgments}

J. RR. is grateful to David Zueco for his encouragement to pursue this topic and for the inspiring discussions that followed. 
V.R. acknowledges the financial support provided by the Villanova University Summer Grant program during the summer of 2026. This work was supported by the Cluster of Excellence Advanced Imaging of Matter (AIM). We acknowledge support from the Max Planck-New York City Center for Non-Equilibrium Quantum Phenomena. The Flatiron Institute is a division of the Simons Foundation.

\end{acknowledgments}

\bibliography{references}

\clearpage
\pagebreak


\appendix

\section{Exact Solution of the CM Hamiltonian and Landau Polaritons}\label{Polariton Solution}
In this section we show that $H_{\textrm{cm}}$ can be solved analytically. To proceed we expand the covariant kinetic term
\begin{equation}
    H_{\textrm{cm}}=\frac{\bm{\Pi}^2}{2m}+\frac{e\sqrt{N}}{m}\mathbf{A}\cdot\bm{\Pi}+ \underbrace{\frac{e^2N\mathbf{A}^2}{2m}+\hbar\omega\left(a^{\dagger}a +\frac{1}{2}\right)}_{H_p}
\end{equation}
For the description of the photon operators we will introduce the displacement coordinate $q$ and its conjugate momentum $\partial_q$ as $a=\frac{1}{\sqrt{2}}\left(q+\partial/\partial q\right)$ and $a^{\dagger}$ defined by conjugation~\cite{spohn2004, GriffithsQM}. The part $H_p$ can be brought to diagonal form by the scaling transformation on the photonic displacement coordinate
\begin{equation}
    u=q\sqrt{\frac{\widetilde{\omega}}{\omega}}\;\; \textrm{where} \;\;\widetilde{\omega}=\sqrt{\omega^2+\omega^2_d}
\end{equation}
with $\omega_d=\sqrt{e^2N/\epsilon_0m \mathcal{V}}$ is the diamagnetic frequency depending on the electron density in the effective mode volume. After this transformation the CM Hamiltonian is
\begin{eqnarray}
   H_{\textrm{cm}}=\frac{\bm{\Pi}^2}{2m}+\frac{e\sqrt{N}}{m}\mathbf{A}\cdot \bm{\Pi}+\frac{\hbar\widetilde{\omega}}{2}\left(-\frac{\partial^2}{\partial u^2}+u^2\right),
\end{eqnarray}
where the quantized field is now 
\begin{equation}
    \mathbf{A}=\sqrt{\frac{\hbar}{\epsilon_0V\widetilde{\omega}}}\mathbf{e}_xu.
\end{equation}
In the Landau gauge the Hamiltonian has translational invariance along the $X$ coordinate which implies that the eigenfunctions in $X$ are plane waves $e^{\textrm{i}K_xX}$. We apply $H_{\textrm{cm}}$ on the plane wave and we have
\begin{eqnarray}
   H_{\textrm{cm}}&=&-\frac{\hbar^2}{2m}\frac{\partial^2}{\partial Y^2}+\frac{m\omega^2_c}{2}\left(Y+\frac{\hbar K_x}{eB}\right)^2\\
   &-&geB u\left(Y+\frac{\hbar K_x}{eB}\right)+\frac{\hbar\widetilde{\omega}}{2}\left(-\frac{\partial^2}{\partial u^2}+u^2\right)\nonumber
\end{eqnarray}
where we also introduced the coupling constant $g=\omega_d\sqrt{\hbar/m\widetilde{\omega}}$. As a next step we define the coordinate
\begin{equation}\label{Y bar}
    \bar Y=Y+\frac{\hbar K_x}{eB}
\end{equation}
and the Hamiltonian simplifies further
\begin{eqnarray}
   H_{\textrm{cm}}=-\frac{\hbar^2}{2m}\frac{\partial^2}{\partial \bar Y^2}+\frac{m\omega^2_c}{2}\bar Y^2-geB u \bar Y+\frac{\hbar\widetilde{\omega}}{2}\left[-\frac{\partial^2}{\partial u^2}+u^2\right]\nonumber
\end{eqnarray}
The Hamiltonian consists of two coupled harmonic oscillators. It is convenient to perform another scaling transformation on $\bar Y$ and $u$ 
\begin{equation}\label{Vpm}
    V_-=-u\sqrt{\frac{\hbar}{\widetilde{\omega}}}\;\; \textrm{and}\;\; V_+=\sqrt{m}\bar Y.
\end{equation}
such that we have both harmonic oscillators in the form of having mass equal to 1. The Hamiltonian then becomes
\begin{equation}
     H_{\textrm{cm}}=-\frac{\hbar^2}{2}\sum_{l=\pm}\frac{\partial^2}{\partial V^2_{l}}+\frac{1}{2}\sum_{l,j=\pm}W_{lj}V_{ l}V_{ j}, 
\end{equation}
where the matrix $W$ is real and symmetric,
\begin{equation}
   W=\left(\begin{tabular}{ c c c c }
		$\omega^2_c$ & $\omega_d\omega_c $ \\
	    $\omega_d\omega_c $ &$\widetilde{\omega}^2$ \end{tabular}\right)
\end{equation}
and as a consequence can be diagonalized by the orthogonal matrix $O$~\cite{faisal1987}, 
\begin{eqnarray}\label{Omatrix}
 &&O=\left(\begin{tabular}{ c c c c }
	    $\frac{1}{\sqrt{1+\Lambda^2}} $ &$\frac{\Lambda}{\sqrt{1+\Lambda^2}}$ \\$-\frac{\Lambda}{\sqrt{1+\Lambda^2}}$ & $\frac{1}{\sqrt{1+\Lambda^2}} $ 
	\end{tabular}\right)\nonumber
\end{eqnarray}
where $\Lambda=\alpha-\sqrt{1+\alpha^2}$ and $\alpha=\frac{\omega^2_c-\widetilde{\omega}^2}{2\omega_d\omega_c}$. The eigenvalues of the matrix $W$ give the new normal modes of the interacting light-matter system. We find them to be
\begin{eqnarray}
 \Omega^2_{\pm}&=&\frac{1}{2}\left(\widetilde{\omega}^2+\omega^2_c\pm\sqrt{4\omega^2_d\omega^2_c+(\widetilde{\omega}^2-\omega^2_c)^2}\right).
\end{eqnarray} 
The Hamiltonian after the orthogonal transformation takes the canonical form
\begin{eqnarray}
 H_{\textrm{cm}}=-\frac{\hbar^2}{2}\sum_{l=\pm}\frac{\partial^2}{\partial S^2_{ l}}+\frac{1}{2}\sum_{l=\pm}\Omega^2_{l}S^2_{ l}.
\end{eqnarray}
The new coordinates $S_{l}$ and conjugate momenta $\partial_{S_{l}}$ are related to the old ones $\{V_{l},\partial_{V_{l}}\}$ through the orthogonal matrix $O$,
\begin{equation}\label{SandV}
    S_{l}=\sum_{j=\pm}O_{jl}V_{j} \; \textrm{and}\; \frac{\partial}{\partial S_{ l}}=\sum_{j=\pm}O_{jl}\frac{\partial}{\partial V_{ j}}.
\end{equation}
Due to the fact that the matrix $O$ is orthogonal the canonical commutation relations are satisfied which implies that we have two independent harmonic oscillators~\cite{faisal1987}. Thus, the eigenfunctions of the interacting system are Hermite functions $\Phi$ of the coordinates $S_{+}$ and $S_{-}$. The full set of eigenfunctions of the system is 
\begin{equation}
    \Psi_{K_x,n_+,n_-}(X,S_{+},S_{-})=e^{\textrm{i}K_xX}\Phi_{n_+}(S_{ +})\Phi_{n_-}(S_{ -})
\end{equation}
with eigenspectrum
\begin{equation}
    E_{n_+,n_-}=\hbar\Omega_+\left(n_+ +\frac{1}{2}\right)+\hbar\Omega_-\left(n_- +\frac{1}{2}\right).
\end{equation}
The frequencies $\Omega_+$ (upper) and $\Omega_-$ (lower) are the two collective Landau polariton modes of the quantum Hall system in the cavity. For completeness, we note that the solution of the polaritons for the CM can be equivalently written in terms of annihilation $b_{\pm}$ and creation $b^{\dagger}_{\pm}$ operators for the polariton quasiparticles. In this representation $H_{\rm cm}$ is written as 
\begin{eqnarray}
    H_{\rm cm }=\hbar \Omega_+\left(b^{\dagger}_{+}b_++\frac{1}{2}\right)+ \hbar \Omega_-\left(b^{\dagger}_{-}b_-+\frac{1}{2}\right)
\end{eqnarray}
with the polariton operators defined $b_{\pm}=S_{\pm}\sqrt{\frac{\Omega_{\pm}}{2}} +\sqrt{\frac{1}{2\Omega_{\pm}}}\partial_{S_{\pm}}$~\cite{GriffithsQM}. It is worth to notice that in the limit $\omega \rightarrow 0$ the lower polariton frequency goes to zero, $\Omega_-\rightarrow 0$, which means that the system becomes gapless.

\section{Zero temperature transport}\label{Transport}

In this section we derive the transport properties of the light-matter system at zero temperature. As we already showed, the Hamiltonian of our system can be written as a sum of a CM and relative part $H=H_{\rm cm}+H_{\rm rel}$. To proceed, we express the eigenstates of $H_{\rm cm}$ as $|\Phi_n\rangle$ and the eigenstates of $H_{\rm rel}$ as $|F_{I}\rangle$ such that it holds
\begin{eqnarray}
H_{\rm cm}|\Phi_{n}\rangle=E_n|\Phi_n\rangle \;\; \textrm{and}\;\; H_{\rm rel}|F_{I}\rangle=E_{I}|F_{I}\rangle
\end{eqnarray}
Then, the eigenstates of the full Hamiltonian $H$ are
\begin{equation}
    |\Psi_{nI}\rangle=|\Phi_n\rangle\otimes |F_{I}\rangle,
\end{equation}
and the full eigenspectrum is $E_{nI}=E_{n}+E_{I}$. 

The Kubo formula for the optical conductivity of the system is~\cite{kubo, ALLEN2006165}
\begin{eqnarray}
 \sigma_{ab}(w)=\frac{\textrm{i}}{w+\textrm{i}\delta}\left(\frac{e^2n_e}{m}\delta_{ab}+\frac{\chi_{ab}(w)}{A}\right)\;\;\; \delta \rightarrow 0^+
\end{eqnarray}
where $a,b=x,y,z$. The first term in the optical conductivity is the Drude term, while the second term is the current-current correlator in the frequency domain, which is defined as the Fourier transform of current-current correlator in the time domain
\begin{eqnarray}
     \chi_{ab}(t)=\frac{-\textrm{i}\Theta(t)}{\hbar }\langle[J_{a}(t),J_{b}]\rangle,
 \end{eqnarray}
with the current operators considered in the interaction picture $\mathbf{J}(t)=e^{\textrm{i}Ht/\hbar}\mathbf{J}e^{-\textrm{i}Ht/\hbar}$~\cite{kubo}. 

Next, we derive the expression of the current operator in CM and relative coordinates.
The expression for the current operator can be obtained by computing the velocity operator of the electrons through the Heisenberg equation of motion~\cite{Landau}
\begin{equation}
 \mathbf{v}_i=\frac{d \mathbf{r}_i}{dt}=\frac{\textrm{i}}{\hbar} [H,\mathbf{r}_i]=\frac{1}{m}\left(-\textrm{i}\hbar \nabla_i-e\mathbf{A}_{\textrm{ext}}(\mathbf{r}_i)-e\mathbf{A}\right).
\end{equation}
Then, the full gauge-invariant current operator is~\cite{Landau}
\begin{eqnarray}\label{Current Operator}
    \bi{J}=e\sum^N_{i=1}\mathbf{v}_i=-\frac{\textrm{i}e\hbar}{m}\sum^N_{j=1}\nabla_j-\frac{e^2N}{m}\bi{A}-\frac{e^2}{m}\sum^N_{i=1}\bi{A}_{\textrm{ext}}(\mathbf{r}_i).\nonumber\\
\end{eqnarray}
Using the expressions derived in Sec~\ref{Polariton Solution} for all the relevant operators we find for the current operator
\begin{equation}\label{Current COM}
    \mathbf{J}=\sqrt{N}\left[-\frac{\textrm{i}e\hbar}{m}\nabla_{\mathbf{R}}-\frac{e^2}{m}\sqrt{N}\mathbf{A}-\frac{e^2}{m}\mathbf{A}_{\textrm{ext}}(\mathbf{R})\right]\equiv \mathbf{J}_{\rm cm}.
\end{equation} 
The above result shows that the total current in the system is equal essentially to current of the CM and depends only on CM related operators. This property has the following important implication
\begin{equation}
    \langle \Psi_{nI}|\mathbf{J}|\Psi_{mJ}\rangle=\delta_{IJ} \langle \Phi_n|\mathbf{J}|\Phi_m\rangle,
\end{equation}

To proceed further we need the expressions for the current operator components in the polaritonic basis.
For the polaritons, we have the CM eigenstates $e^{\textrm{i}K_xX}\phi_{n_+}(S_+)\phi_{n_-}(S_-)\equiv |K_xn_+n_-\rangle$ and the eigenergies $E_{n_+n_-}=\hbar\Omega_+\left(n_++\frac{1}{2}\right) + \hbar\Omega_-\left(n_-+\frac{1}{2}\right)$.
The $x$ and $y$ components of the current operator in terms of the polaritonic coordinates ${S_\pm}$ are
\begin{widetext}
    \begin{eqnarray}
 J_x&=&\frac{e^2\sqrt{N}B}{m^{3/2}}\left[\frac{\sqrt{m}}{eB}\left(-\textrm{i}\hbar\nabla_X-\hbar K_x\right)+\frac{S_+(1-\eta\Lambda)+S_-(\Lambda+\eta)}{\sqrt{1+\Lambda^2}}\right]\nonumber\\
 J_y&=&-\frac{\textrm{i}e\hbar}{m}\sum^N_{j=1}\partial_{y_j}=\frac{-\textrm{i}e\hbar}{\sqrt{m}}\sqrt{\frac{N}{1+\Lambda^2}}\left[\partial_{S_+}+\Lambda\partial_{S_-}\right].
\end{eqnarray}
\end{widetext}
Moreover, the current operators can be written using the polaritonic annihilation and creation operators as follows
\begin{widetext}
    \begin{eqnarray}
    J_x&=&\frac{e^2\sqrt{N}B}{m^{3/2}}\sqrt{\frac{\hbar}{2(1+\Lambda^2)}}\left[\frac{\sqrt{m}}{eB}\left(-\textrm{i}\hbar\nabla_X-\hbar K_x\right)+\frac{\Lambda+\eta}{\sqrt{\Omega_-}}\left(b^{\dagger}_-+b_-\right)+\frac{1-\eta\Lambda}{\sqrt{\Omega_+}}\left(b^{\dagger}_++ b_+\right)\right]\\
J_y&=&-\textrm{i}e\sqrt{\frac{ \hbar N}{2m(1+\Lambda^2)}}\left[\sqrt{\Omega_+}\left(b_+-b^{\dagger}_+\right)+\Lambda\sqrt{\Omega_-}\left(b_--b^{\dagger}_-\right)\right]
\end{eqnarray}
\end{widetext}
From the above we can obtain the matrix representation of the current operator on the polariton basis
\begin{widetext}
       \begin{eqnarray}
\langle n_+n_-K^{\prime}_x|J_x|K_xm_+m_-\rangle&=&\frac{e^2\sqrt{N}B}{m^{3/2}}\sqrt{\frac{\hbar}{2(1+\Lambda^2)}}\Bigg[\frac{\Lambda+\eta}{\sqrt{\Omega_-}}\delta_{n_+m_+}\left(\sqrt{m_-+1}\delta_{n_-,m_-+1}+\sqrt{m_-}\delta_{n_-,m_--1}\right)\nonumber\\
    &+&\frac{1-\eta\Lambda}{\sqrt{\Omega_+}}\delta_{n_-m_-}\left(\sqrt{m_+}\delta_{n_+,m_+-1}+\sqrt{m_++1}\delta_{n_+,m_++1}\right)\Bigg]\delta_{K^{\prime}_xK_x}\\
\langle n_+n_-K^{\prime}_x|J_y|K_xm_+m_-\rangle&=&-\textrm{i}e\sqrt{\frac{ \hbar N}{2m(1+\Lambda^2)}}\Big[\sqrt{\Omega_+}\delta_{n_-m_-}\left(\sqrt{m_+}\delta_{n_+,m_+-1}-\sqrt{m_++1}\delta_{n_+,m_++1}\right)\nonumber\\
    &+&\Lambda\sqrt{\Omega_-}\delta_{n_+m_+}\left(\sqrt{m_-}\delta_{n_-,m_--1}-\sqrt{m_-+1}\delta_{n_-,m_-+1}\right)\Big]\delta_{K^{\prime}_xK_x}
\end{eqnarray} 
\end{widetext}
The current operators are diagonal with respect to the plane-wave states $e^{\textrm{i}K_xX}$. With this, we can finally obtain expressions for the current-current correlators from their zero-temperature Lehmann representation \cite[Chap. 7]{altland2010condensed}
\begin{eqnarray}
    \chi_{ab}(w)&=&\sum_{m_+,m_-} \frac{\langle 00|J_{a}|m_+m_-\rangle\langle m_+m_-|J_{b}|00\rangle}{w+(E_{00}-E_{m_+m_-})/\hbar +\textrm{i}\delta}\nonumber\\
    &-& (00 \leftrightarrow m_+m_-).
\end{eqnarray}
The current operators are linear in the polaritonic annihilation and creation operators and thus allow only for single-polariton transitions to occur, which implies that in the denominator of the response function only single polariton energies show up $\Omega_{\pm}$. Using the formulas for the matrix representation of the components of the current operator we find the following analytically exact expressions for the correlators
\begin{widetext}
    \begin{eqnarray}
 \chi_{xx}(w) &=& -\frac{\hbar N e^4 B^2}{(1+\Lambda^2) m^3}\Big[\frac{(1 - \eta \Lambda)^2}{2 \Omega_+} \left(\frac{1}{w+\Omega_++\textrm{i}\delta}-\frac{1}{w-\Omega_++\textrm{i}\delta}\right) + \frac{(\Lambda + \eta)^2}{2 \Omega_-} \left( \frac{1}{w+\Omega_-+\textrm{i}\delta}-\frac{1}{w-\Omega_-+\textrm{i}\delta} \right) \Big],\nonumber\\
 \chi_{xy}(w)&=&\frac{\hbar Ne^3B}{(1+\Lambda^2)m^2}\Big[ \Lambda(\Lambda+\eta)\frac{\textrm{i}}{2}\left(\frac{1}{w+\Omega_-+\textrm{i}\delta}+\frac{1}{w-\Omega_-+\textrm{i}\delta}\right)
 +(1-\eta\Lambda)\frac{\textrm{i}}{2}\left(\frac{1}{w+\Omega_++\textrm{i}\delta}+\frac{1}{w-\Omega_++\textrm{i}\delta}\right)\Big],\nonumber\\
\chi_{yy}(w)&=&-\frac{\hbar Ne^2}{(1+\Lambda^2)m}\left[\frac{\Omega_+}{2}\left(\frac{1}{w+\Omega_++\textrm{i}\delta}-\frac{1}{w-\Omega_++\textrm{i}\delta}\right)+\frac{\Lambda^2\Omega_-}{2}\left(\frac{1}{w+\Omega_-+\textrm{i}\delta}-\frac{1}{w-\Omega_-+\textrm{i}\delta}\right) \right].
\end{eqnarray}
\end{widetext}
With the above results and using the Kubo formula, the expressions for the optical and the CD conductivities can be straightforwardly obtained.
\end{document}